\documentclass[12pt]{article}
\usepackage{psfig}
\begin{document}

\title{Renormalization study of two-dimensional convergent solutions of the porous medium
equation}
\author{S. I. Betel\'{u}, D. G.
Aronson\\ School of Mathematics,
University of Minnesota,\\
 Minneapolis, Minnesota 55455\\
\\
S. B. Angenent\\
Mathematics Department, University of Wisconsin,\\
 Madison, Wisconsin
53706}
\maketitle

\begin{abstract}
In the focusing problem we study a solution of the porous medium equation 
$u_t=\Delta \left( u^m\right)$ whose initial distribution is positive in the
exterior of a closed non-circular two dimensional region, and zero inside. 
We implement a numerical scheme that renormalizes the solution each time
that the average size of the empty region reduces by a half. The initial
condition is a function
with circular level sets distorted with a small sinusoidal perturbation of
wave number $k\geq 3$. We find that for nonlinearity exponents $m$ smaller
than a critical value which depends on $k$, the solution tends to a
self-similar regime, characterized by rounded polygonal interfaces and
similarity exponents that depend on $m$ and on the discrete rotational
symmetry number $k$. For $m$ greater than the critical value, the final
form of the interface is circular.
\end{abstract}

\section{Introduction}
  
This work deals with solutions to the nonlinear diffusion equation
\begin{equation}
\frac{\partial u}{\partial t}=\Delta \left( u^m\right) ,  \label{ecuacion}
\end{equation}
where $m>1$ is constant and $\Delta $ denotes the Laplace operator in $%
\mathbf{R}^d$. Equation (\ref{ecuacion}) arises in various problems, such as the      
spreading of viscous gravity currents \cite{Hupp82,Diez92}, the diffusion of strong
thermal waves \cite{Baren79}, and the isentropic flow of an ideal gas in a
homogeneous 
porous medium \cite{Aro86,Baren52}. In the latter application, $u$ represents the scaled   
density of the gas. For $m=1$ equation (\ref{ecuacion}) is the classical heat
conduction
equation and there are other applications which involve values of $m<1$.
One
of the salient features of the ``slow diffusion'' case $m>1$ is the  
occurrence of interfaces moving with finite velocity, that separate empty  
regions where $u=0$ from regions where $u>0$. More detailed information
about the properties of solutions to equation (\ref{ecuacion}) can be found in \cite{Aro86}.

In order to discuss the behavior of solutions to equation (\ref{ecuacion}) it is
convenient to replace the independent variable $u$ with
\[
v=\frac m{m-1}u^{m-1}.
\]
In the gas flow interpretation, $v$ represents the scaled pressure via the
ideal gas law. If $u$ satisfies (\ref{ecuacion}) then $v$ satisfies
\begin{equation}
\frac{\partial v}{\partial t}=(m-1)v\Delta v+\left| \nabla v\right| ^2,
\label{pressure}
\end{equation}
where $\nabla $ is the gradient operator in\textbf{\ }$\mathbf{R}^d$.
  
Here we are interested in the \textit{focusing problem }in which we solve
equation (\ref{pressure}) starting from an initial distribution which is positive in
the
\textit{exterior} of a closed bounded region $D$ and zero inside $D$ (see
Fig. \ref{sketch}).
At
some finite time $T>0$, called the \textit{focusing time}, $v$ will for
the
first time be positive throughout the initially empty region $D$.

In \cite{Aro93} it is shown that for each $m\in \left( 1,\infty \right) $ there is
a
unique similarity exponent $\beta (m,d)\in \left( \frac 12,1\right) $ and
a
one-parameter family of functions $F_c$ such that the functions
\begin{equation}
v_c(x,t)=(T-t)^{2\beta -1}F_c\left( \frac{\left| x\right| }{(T-t)^\beta }%
\right) .  \label{3}
\end{equation}
with $\beta =\beta \left( m,d\right) $, are self-similar solutions to (\ref{pressure}).
Moreover, there exists a $\gamma (d,m)$ such that
\[
F_c(\xi )\left\{
\begin{tabular}{l}
$=0$ for $0\leq \xi \leq \left( \frac c\gamma \right) ^\beta $ \\
$>0$ for $\xi >\left( \frac c\gamma \right) ^\beta $%
\end{tabular}
\right. ,
\]
where $\gamma =\gamma (d,m)$. Thus the $v_c(x,t)$ are focusing solutions
to
equation (\ref{pressure}) with interfaces given by
\[
\frac{\left| x\right| }{(T-t)^\beta }=\left( \frac c\gamma \right) ^\beta
.
\]

The focusing problem is well studied in the radially symmetric case. In
particular, it is proven in \cite{Ange95,Ange96} that some member of the Graveleau family
(\ref{3}) describes locally, to leading order, the focusing behavior of
essentially any radially symmetric focusing solution to the pressure
equation (\ref{pressure}). The situation is much more complicated in the absence of
radial symmetry. Physical experiments involving convergent gravity flows
($%
m=4,d=2$) followed by numerical experiments \cite{Diez98} indicate that both large
and
small deviations from rotational symmetry may be amplified as the flow tries
to
fill the hole. A formal linear stability analysis for general $m$ and $d$
(cf. \cite{Ange97} and Appendix 2) shows that the Graveleau solutions are indeed
unstable, at least when $m$ is close to unity, and that the number of
unstable modes increases as $m\searrow 1$. This suggests that a sequence
of
bifurcations occurs as $m$ decreases from $\infty $ to $1$. The existence
of
these bifurcations is proved in \cite{Ange97}. The bifurcating solutions are
self-similar, but not axially symmetric. They are invariant with respect
to the action of the group $O(d-1,\bf{R})$ of $(d-1)\times(d-1)$ real
orthogonal matrices. In this paper we restrict our attention to the plane
case $d=2$. In this case the results of \cite{Ange97} show that for
 any $\bar m\in (1,\infty )$ there exists an integer $
k_{*}(\bar m)\in (2,\infty )$ such that a symmetry breaking bifurcation
from the
radially symmetric Graveleau solutions occurs at some $m_k\in (1,\bar m)$
for
each $k>k_{*}(\bar m)$. The bifurcating solutions are non-radial
self-similar
focusing solutions possessing $k$-fold symmetry, i.e., the symmetry of
$\cos
(k\theta )$. Numerical studies show that for each $k\geq 3$ the
bifurcation
point $m=m_k$ is unique so that the Graveleau solutions are linearly
stable 
with respect to perturbations with $k$-fold symmetry for $m>m_k$ and
unstable for $m_k>m>1$. Moreover, the $m_k$ are ordered
\[
\infty >m_3>m_4>...>m_k>...\searrow 1.
\]
Crude estimates for the first four $m_k$ are given in Table I (See also Table IV in
Appendix 2).
\[
\begin{tabular}{ll}
k & estimate \\
3 & $m_3\in (1.69,1.7)$ \\
4 & $m_4\in (1.32,1.321)$ \\
5 & $m_5\in (1.18,1.19)$ \\
6 & $m_6\in (1.12,1.13)$%
\end{tabular}
\]
\begin{center}
\textbf {Table I:} Estimates of the values of $m_k$ for which the
circular
self-similar 
solution bifurcates into a non-circular solution. 
\end{center}

The results in \cite{Ange97} give no information about stability with respect to
perturbations with wave number $k=2$. The experimental and numerical
results
in \cite{Diez98} suggest instability. This is confirmed by numerical linear
stability
analysis which shows that the Graveleau interface is unstable with respect
to perturbations with wave number 2 for all values of $m$. In particular,   
there appear to be no bifurcating branches of self-similar focusing
solution
with the 2-fold symmetry. We discuss this briefly in Section 5 of this
paper
and in more detail in reference \cite{Betelu99}.

In this paper we carry out detailed numerical studies of focusing
solutions
to equation (\ref{pressure}) whose initial distributions have interfaces which are
circles with small perturbations. Mainly, we deal with initial conditions whose interfaces
are of the form
\begin{equation}
r=a\left(1+\varepsilon \cos (k\theta ) \right) \label{initiali}
\end{equation}
with $\varepsilon \ll a$, although in Section 5 we will consider
perturbations with mixed modes. By using a numerical renormalization
technique inspired by the pioneering work of Chen and Goldenfeld \cite{Chen,Golden}
we
are able to follow the evolution of the interface to times very close to
the
focusing time. Thus we are able to obtain very detailed information about
the asymptotic form of the solution as it focuses. As we shall see below,
for single mode perturbations with $k$-fold symmetry, the numerical
results
indicate that the leading term in the focusing asymptotics is a
self-similar
solution of the form
\begin{equation}
v=(T-t)^{2\delta -1}V_c\left( \frac r{(T-t)^\delta },\theta \right)
\label{4}
\end{equation} 
where $c$ is a parameter, $\delta =\delta \left( k,m\right) $ is the
similarity exponent, and $V_c$ satisfies
\begin{equation}
V_c(\zeta ,\theta )=V_c(\zeta ,\theta +\frac{2\pi n}k)\mbox{ for }
n=1,...,k-1. \label{periodicity}
\end{equation}
Moreover, the focusing interface is asymptotically of the form
\[
r=(T-t)^\delta A\left( \theta \right) ,
\]
and $\delta \left( k,m_k\right) =\beta \left( 2,m_k\right) $, where $\beta
$
is the similarity exponent for the Graveleau solution and $m_k$ is the
bifurcation point found in \cite{Ange97}. Therefore we conclude that the functions
given by (\ref{4}) are the bifurcating solutions found in \cite{Ange97}.

\section{Numerical scheme and renormalization procedure}

In order to solve Eq. (\ref{pressure}) numerically, we discretize a circular
domain $D=[0,R]\times [0,2\pi /k]$  with a uniform polar
grid of interval sizes $\delta r=R/N_r$ and $\delta \theta =2\pi /N_\theta $.  The
numerical solution is stored in a matrix as $v(r,\theta,t )=v_{ij}$ with
$r=i\delta
r$ and 
$\theta =j\delta \theta$.
In order to integrate the PDE in time, we use an explicit Euler scheme
\begin{equation}
v_{ij}(t+\delta t) = v_{ij}(t) + \delta t H_{ij}   \label{euler}
\end{equation}
where $H_{ij}$ is a finite differences approximation of the derivatives of the right hand
side of Eq. (\ref{pressure}). We
compute the derivatives of the term $\left| \nabla v\right|
^2=v_r^2+v_\theta ^2/r^2$ with a second order upwind ENO (essentialy non-oscillatory)
scheme \cite{Osher91,Shu88} (see Appendix 1). This method
guarantees that the numerical scheme will be able to describe accurately
the discontinuities on the first derivative that spontaneously appear in
the case $m=1$ (Hamilton-Jacobi limit) and at the interface. Due to the diffusive nature of the
Laplacian term in Eq. (\ref{pressure}), the corresponding derivatives are computed with standard
centered second order approximations, 
\begin{equation}
\Delta v=v_{rr}+\frac{v_r}r+\frac{v_{\theta \theta }}{r^2} \simeq 
\end{equation}
\begin{equation}
\simeq \frac{v_{i+1,j}-2v_{ij}+v_{i-1,j}}{\delta r^2}+\frac{%
v_{i+1,j}-v_{i-1,j}}{2\delta r\,r_i}+\frac{v_{i,j+1}-2v_{ij}+v_{i,j-1}}{%
\delta \theta ^2 r_i^2}.  \label{lapnum}
\end{equation}

Near the interface there is a discontinuity in the first derivative that may
generate numerical errors when we compute the Laplacian with Eq. (\ref
{lapnum}). Therefore, for the grid points $v_{ij}$ situated at a distance smaller than
$2\delta r$ from the interface, the Laplacian is computed by linearly 
extrapolating in the variable $r$ the values of the Laplacian from
the nodes of behind, which were previously computed with Eq. (\ref{lapnum}).

At the boundaries $\theta =\pm \pi /k$ we apply the boundary condition of
periodicity $v(-\pi /k)=v(\pi /k)$. This is simply a convenience, since 
computations without this forced symmetry yield equivalent results.
At the boundary $r=R$ we apply the
boundary condition $v_{rr}=0$. This is equivalent to a first order linear
extrapolation of the ghost points $v_{N_r+1,j}$ outside $R$, which are needed
in order to compute the derivatives at $R$. Finally, at $r=0$ we set $v=0$.

We start the integration with a rather arbitrary initial condition, 
\begin{equation}
v^{(0)}(r,\theta ,t_0)=r-a_0\left( 1+\varepsilon \cos k\theta \right)
,\hspace{1.5cm} \varepsilon =0.1  \label{initial}
\end{equation}
which describes a function whose contour lines are perturbed circles, and
which interface is given by Eq. (\ref{initiali}). The
exact form of the initial condition is not very critical, because the
asymptotic solution only depends on $k$ (which determines the symmetry) 
and $m$, as was verified numerically.

We integrate the diffusion equation over a sequence of time intervals
$(t_n,t_{n+1})$ starting with $n=0$, and renormalize the
solution at each right hand end point $t=t_{n+1}$ before continuing the
integration. The renormalization times $t_{n+1}$ are taken to be the times
when the average radius of the interface\footnote{
The interface $a(\theta ,t)$ is defined as $v(a(\theta
,t),\theta ,t)=0$. In order to avoid numerical problems due to the
 numerical diffusion near the interface, we extrapolate the positive values 
of $v$ up to $v=0$.} 
\begin{equation}
\overline{a}(t)=\frac 1{2\pi }\int_0^{2\pi }a(\theta ,t)d\theta
\end{equation}
reaches half of its initial value. The renormalized solution is defined to
be
\begin{equation}
v^{(n+1)}(r,\theta ,t_{n+1})= Z^{(n)}\cdot v^{(n)}(r/2,\theta
,t_{n+1})  \label{zeta}
\end{equation}
for $n=0,1,2,...$.
This transformation is performed by linearly
interpolating the values of the grid, 
\[v_{ij}^{(n+1)}=Z^{(n)}\cdot
v_{i/2,j}^{(n)} \hspace{1cm} \mbox{ for even }i
\]
and
\[v_{ij}^{(n+1)}=Z^{(n)}\cdot
(v_{i/2,j}^{(n)}+v_{i/2+1,j}^{(n)})/2 \hspace{1cm} \mbox{ for odd }i.
\]
 The constant $Z^{(n)}$
is taken as the reciprocal of the maximum value of the function $v$ in the
renormalized domain of integration 
\begin{equation}
Z^{(n)}=\left( \max_{r\leq R/2}v^{(n)}(r,\theta ,t_{n+1})\right) ^{-1}.
\label{zetadef}
\end{equation}
The superscript $(n)$ indicates how many renormalizations we have made up to the
 time $t_{n+1}$. The errors introduced in the linear interpolation are of second
order, the same as in the discretization of the derivatives.
It is very important to determine the time $t_{n+1}$ accurately, and this 
is done by first detecting the exact time when the interface crosses half 
 of its initial value at the previous renormalization, and then using this
information to re-compute 
the time step $\delta t$ and the solution such that the interface reaches
exactly the half of the initial value.

We summarize the procedure as follows:

\begin{enumerate}
\item  Initialize $v^{(0)}(r,\theta ,t_0)$ with Eq. (\ref{initial}). Let $n=0$.

\item  Starting from $t=t_n$ solve Eq. (\ref{pressure}) with standard finite
 differences until the time $t=t_{n+1}$ where
$\overline{a}(t_{n+1})=\overline{a}(t_n)/2$. Here we have $v^{(n)}(r,\theta ,t_{n+1})$.

\item  Renormalize $v^{(n+1)}(r,\theta ,t_{n+1})= Z^{(n)}\cdot
v^{(n)}(r/2,\theta ,t_{n+1}).$

\item  While 
\begin{equation}
\varepsilon_n =\sqrt{\frac {1}{\pi (R/2)^2} {\int \int }_{r\leq R/2}
\left| v^{(n+1)}(r,\theta ,t_{n+1})-v^{(n)}(r,\theta ,t_n)\right| ^2r\ dr\
d\theta} >\tau  \label{error}
\end{equation}
let $n=n+1$ and return to step 2.
\end{enumerate}
where $\tau$ is a tolerance. In the experiments reported below, we have set
$\tau= 10^{-6}$.

We found that when we repeat this procedure $n$ times, the solution
typically converges for $n\geq 50$. 
This convergence is a necesary condition for the existence of a
self-similar
solution. In Fig. (\ref{convergence}) we show the convergence of the
iterative procedure by plotting the difference of succesive approximations
given by Eq. (\ref{error}) and the departure of $Z^{(n)}$ from its
asymptotic value.

In order to compute the exponent $\delta $ we note that once the scheme
has converged, 
\[
v^{(n+1)}(r,\theta ,t_{n+1})=v^{(n)}(r,\theta ,t_n) 
\]
and 
\[
Z^{(n+1)}=Z^{(n)}=Z
\]
(here we are ignoring the numerical errors). Then, from Eq. (\ref{zeta}), 
\begin{equation}
v^{(n)}(r,\theta ,t_n)=Z\cdot v^{(n)}(r/2,\theta ,t_{n+1}).  \label{converge}
\end{equation}
Thus, the solution at time $t_{n+1}$ is proportional to the solution at time 
$t_n$ with the distances scaled by a factor two.

By substituting $v$ from Eq. (\ref{4}), 
\begin{equation}
\left( -t_n\right) ^{ 2 \delta -1}V\left( \frac r{\left( -t_n\right)
^\delta },\theta \right) =Z\left( -t_{n+1}\right) ^{2\delta -1}V\left(
\frac r{2\left( -t_{n+1}\right) ^\delta},\theta \right)
\end{equation}
(here we have put $T=0$)
it is clear that 
\begin{equation}
t_n=2^{1/\delta} t_{n+1}\hspace{1cm}\mbox{and }\hspace{1cm} \left(
-t_n\right) ^{2
\delta -1}=Z\left(
-t_{n+1}\right) ^{ 2\delta -1}.
\end{equation}
Solving these algebraic equations we find that the similarity
exponent is given by 
\begin{equation}
\delta^{-1} =\log _2\left( \frac 4Z\right) .  \label{alfa}
\end{equation}

\section{Validation}

\begin{figure}

\hspace{3cm}\psfig{file=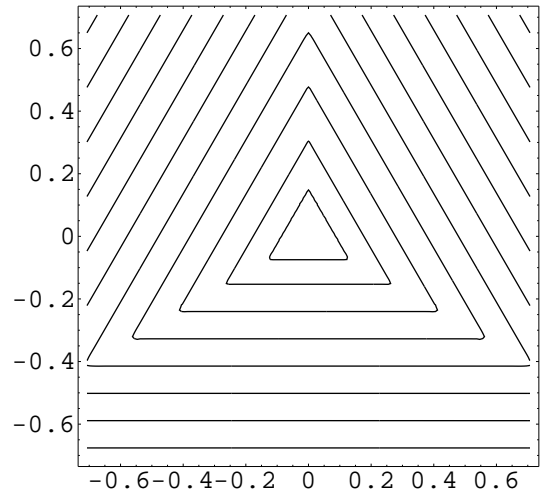}\\

\hspace{3cm}\psfig{file=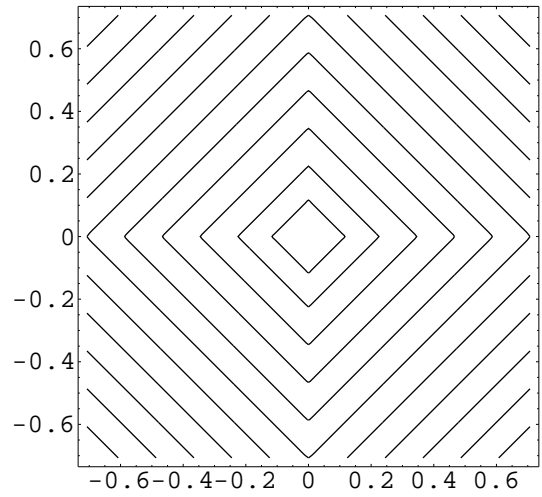}\\

\hspace{3cm}\psfig{file=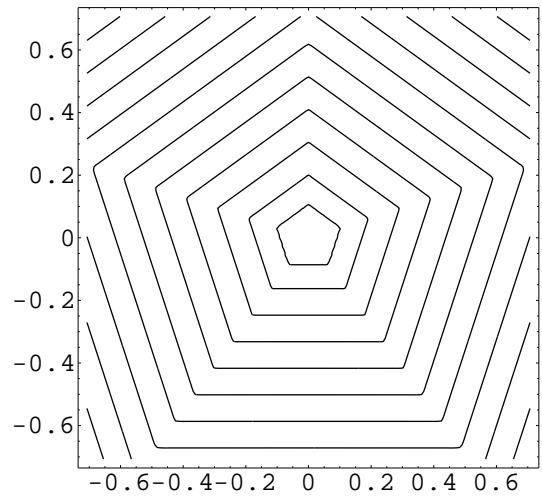}\\

\caption{ Polygonal contour lines of the asymptotic solutions for $m=1$
and $k=3$, 
$4$ and $5$. The innermost contour line is the interface.}
\label{levelsetm1}
\end{figure}

\begin{figure}
\psfig{file=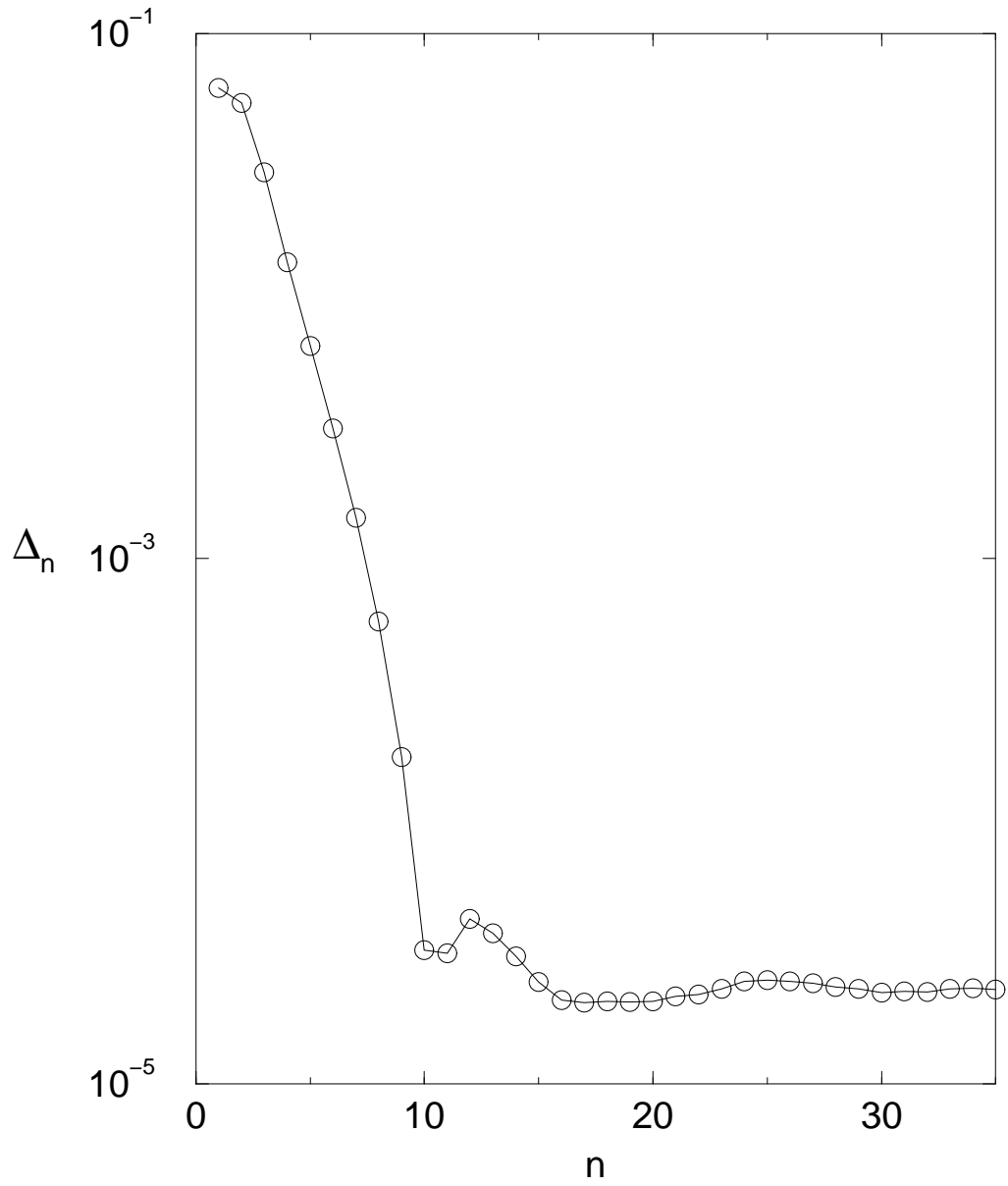}
\caption{Validation with the exact solutions for $m=1$: averaged
differences
between the numerical solution with the exact solution for $k=3$ as 
a function of the number of renormalizations $n$.}\label{validm1}
\end{figure}

\begin{enumerate}
\item  In order to show the accuracy of this numerical computation, we
compare in Table II the numerical solution for the circularly symmetric case
with the Graveleau solution, and we find a good agreement in the 
exponents ($\delta=\beta$ in this case, see Eqs. \ref{3} and \ref{4}) \footnote{We
implemented the numerical scheme in FORTRAN language and we 
made the computations on an IBM RS/6000 computer}.
\[
\begin{array}{ccc}
m & Numeric & ^{\prime }Exact^{\prime } \\ 
1.0 & 1.00000 & 1 \\ 
1.1 & 0.976749 & 0.976744 \\ 
1.2 & 0.956533 & 0.956517 \\ 
1.3 & 0.938768 & 0.938756 \\ 
1.4 & 0.923066 & 0.923071 \\ 
1.5 & 0.909103 & 0.908983 \\ 
1.6 & 0.896595 & 0.896346 \\ 
1.7 & 0.885311 & 0.884940 \\ 
1.8 & 0.875052 & 0.874546 \\ 
1.9 & 0.865675 & 0.865052 \\ 
2.0 & 0.857052 & 0.856333
\end{array}
\]

\begin{center}
\textbf{Table II:} Exponents for the circular case, $R=1, N_r=300$
\end{center}

This comparison checks the accuracy of the renormalization procedure and the
temporal integration.

\item  We also computed the solutions corresponding to the Hamilton-Jacobi
case $m=1$, and we compared the numerical result with the exact solution.
The interfaces of the exact solution are regular polygons of $k$ sides ($%
k\geq 3$), and the function $v$ is a set of $k$ planes of the form 
\begin{equation}
v(r,\theta ,t)=c(x+a(t))=c(r\cos \theta +a(t))\hspace{1cm} -\pi /k\leq \theta
<\pi /k \label{hjexact}
\end{equation}
\[
v(r,\theta +2\pi n/k,t)=v(r,\theta ,t)\qquad for\ n=1\,..\,k-1
\]
where $a(t)=ct$ and $c\ $is a constant . The level lines are regular
polygons and the similarity exponent is $\delta =1$. This solution
exhibits a set of $k$ discontinuities in the derivatives at the apeses of
the polygonal level sets. The level sets of the numerical solution are shown in 
Fig. \ref{levelsetm1} for $k=3, 4,$ and $5$. The numerical solution gives the value of the exponent with an
absolute error
of the order of $10^{-5}$ when a discretization of $N_r=300$ and
 $N_\theta=300$ is used. 
The relative departure of the numerical solution from the theoretical
solution is
\begin{equation}                                                      
\Delta_n=\frac 1{v_{max}}\sqrt{\frac 1{\pi (R/2)^2}
{\int \int }_{r\leq R/2}\left| v^{(n)}(r,\theta ,t_n)-v(r,\theta ,t_n)\right| ^2r\
dr\ d\theta }  \label{error2}
\end{equation}
where $v$ is computed with Eq. (\ref{hjexact}) and $v_{max}$ is the
maximum value of the solution in the
integration domain. The value of $c$ is
obtained from the numerical solution, $c=v_r^{(n)}(a(0,t),0,t)$.
This difference is shown in Fig. (\ref{validm1}) as a
function of the number of renormalizations $n$. Clearly, the numerical
computation converges to the exact solution with relative errors smaller
than $10^{-4}$.

This comparison checks the accuracy of the two-dimensional evaluation of the $\left| \nabla
v\right| ^2$ term and the renormalization routine.

\item  We also validated the code for the case of two colliding plane fronts
($d=1$), where the exact solution is 
\[
v(x,t)=x+t.
\]
The exact value of the exponent is $1$ for all $m$. For $m=2$ the
numerical exponent
are close to $1.0024$, and the level lines depart from the straight
theoretical lines in less than one part in $500$. In this case, the
Laplacian
term is zero, however, since the numerical code is written in polar
coordinates, it serves as a test for the evaluation of the terms in the
Laplacian.
\end{enumerate}

\section{Non-circular self-similar collapse}

We summarize here the results for non-circular collapses, starting with the
initial condition Eq. (\ref{initial}). The numerical solution shows that the
final shape of the interfaces are rounded polygons. The shape of the
interfaces may be characterized by the ratio between the maximum to minimum
radii of the interface 
\[
I=\frac{\min a(\theta ,t)}{\max a(\theta ,t)} \qquad 0\leq \theta <2\pi . 
\]
For instance, for a polygon of $k$ sides, $I$ is equal to $\cos \pi /k$.

The computations were performed in a domain of radius $R=1$ with
$a_0=0.1665$ and 
$\varepsilon =0.1$. The number of grid points in the $r$ direction was $N_r=300$ and the
effective number of points in the $\theta$ interval $(0,2\pi )$, $N_\theta
=300$. Table III shows the exponents that result from this computation.
 The value of $I$ is indicated in parentheses.

\begin{center}
\[
\begin{array}{cccc}
m & k=3 & k=4 & k=5 \\
1.0 & 1.0000\ (0.5029) & 1.0000\;(0.709) & 1.0000\;(0.810) \\
1.1 & 1.0000\ (0.5029) & 1.0000\;(0.709) & 0.9997\;(0.811) \\
1.2 & 1.0000\;(0.5027) & 0.9985\;(0.710) & - \\
1.3 & 1.0000\ (0.5025) & 0.9587\;(0.770) & - \\
1.4 & 0.9996\;(0.5021) & - & - \\
1.5 & 0.9964\;(0.5027) & - & - \\
1.6 & 0.9839\ (0.5075) & - & -
\end{array}
\]
\textbf{Table III:} Exponents for the non-circular collapse
\end{center}

\begin{figure}
\psfig{file=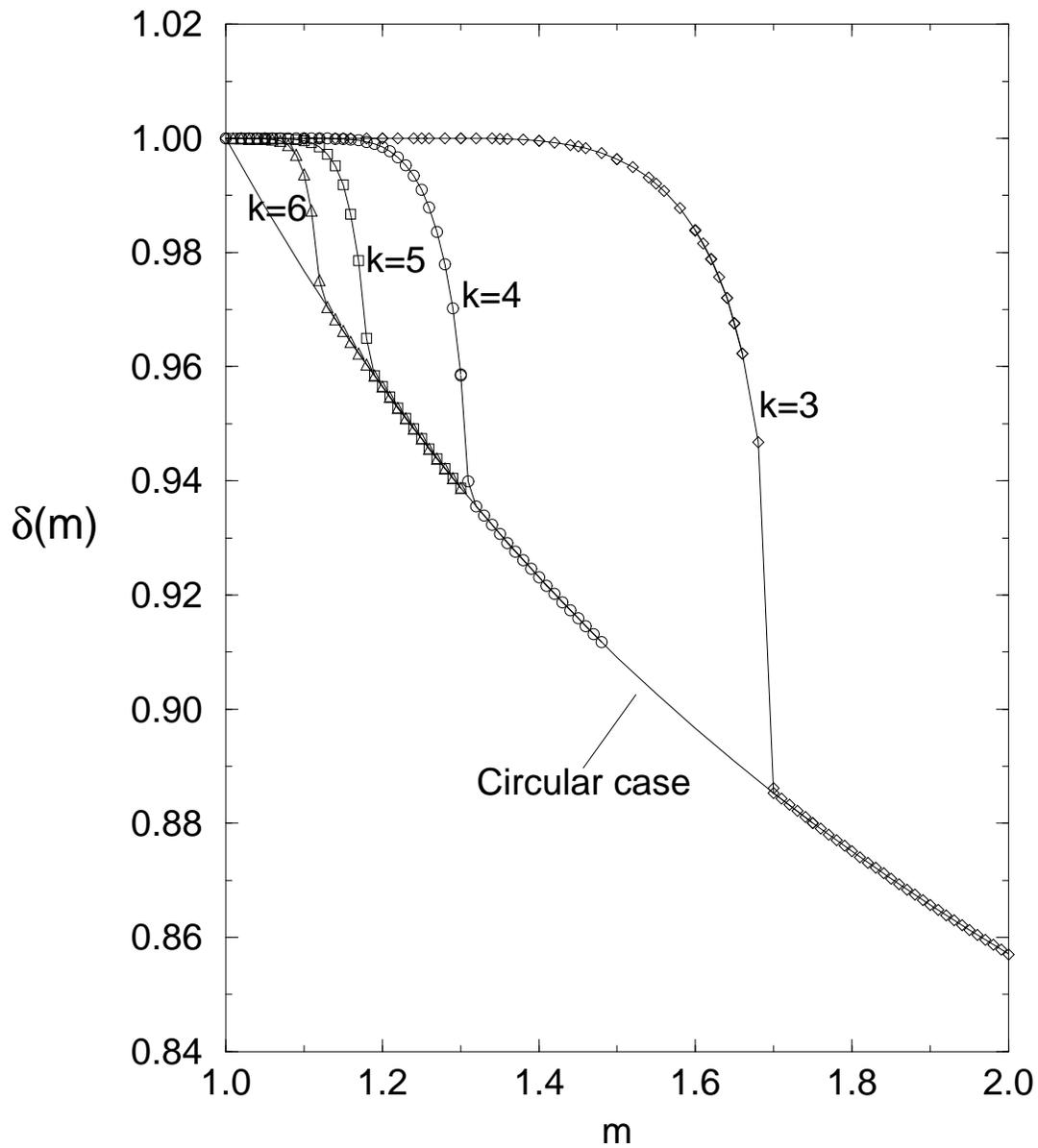}
\caption{  Exponents $\delta $ as a function of the nonlinear exponent
$m$,
for 
$k=3,$ $4$, $5$ and $6$ (dots and line). The line lowermost line 
represents the
Graveleau exponents
$\beta$ for the circular case.}\label{exponents}
\end{figure}

\begin{figure}

\hspace{3cm}\psfig{file=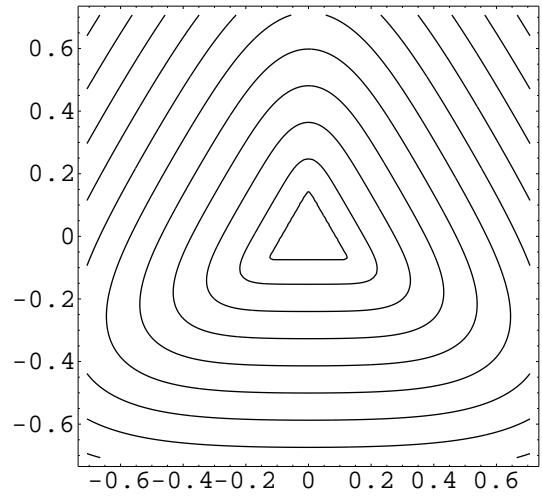}\\

\hspace{3cm}\psfig{file=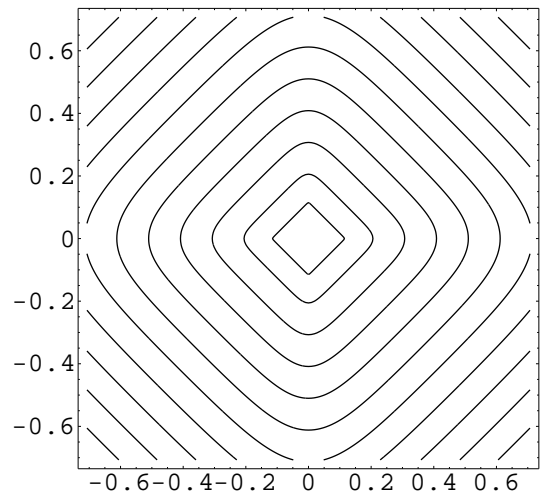}\\

\hspace{3cm}\psfig{file=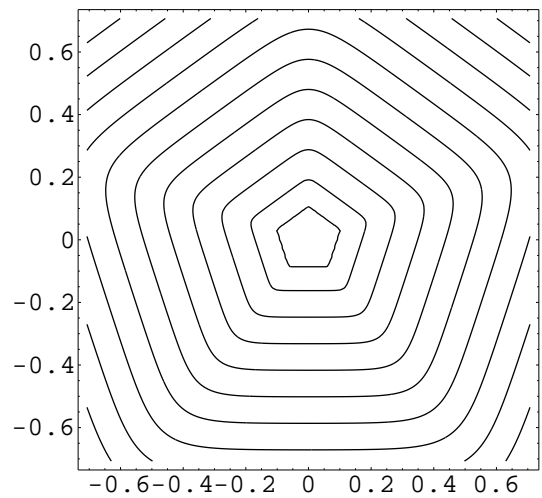}\\

\caption{  Level sets of the self-similar non-circular solutions for
$k=3$,
$4$ and $5$
for
$m=1.5$, $m=1.2$ and $m=1.1$, respectively. The innermost contour
line is the interface.}\label{levelsets}
\end{figure}

The dashes indicate that for this values of $k$ and $m$ the solution tends
to the circular case (Graveleau solution). This result serves as another
validation, since the solution becomes unstable for the values of $m$
reported in Appendix 2 (see also Table I). Note that for $m=1.3$ and
$k=4$, the
aspect ratio
$I$ becomes closer to the circular case $(=1)$ because we are near the
bifurcation. In Fig. (\ref{exponents}) we plot
the exponents for the radially symmetric case, and for the non-circular
cases $k=3$, $4$, $5$ and $6$. Note the continuity of the curves at the
stability limits. The case $k=2$ is not included because the corresponding
evolution is not self-similar.
In Fig. \ref{levelsets} we show typical asymptotic solutions for three different values of $k$.

We observe that the form of the asymptotic solutions only depend on the
value of $k$ that determine the symmetry and $m$, even when the initial
condition has a hole with a more complex shape than the described by Eq.
(\ref{initial}).

\section{Collapse without self-similarity}

The computations shows that self-similar solutions are obtained when the initial 
condition has rotational k-fold symmetry, with $k>2$.
When the initial condition does not satisfy this symmetry condition, then the collapse may not be
self-similar. In \cite{Diez98} the experiments indicate that when the initial shape
is elongated in
one direction, then there is not a self-similar solution. Moreover, the dimensions of the  mayor axis
and the minor axis seem to follow different power laws.

\begin{figure}

\hspace{3cm}\psfig{file=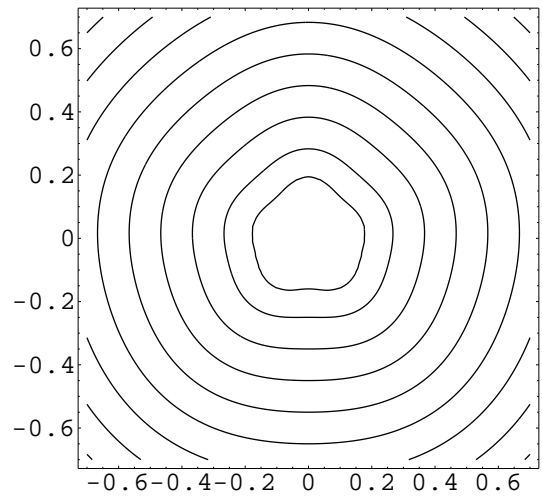}\\

\hspace{3cm}\psfig{file=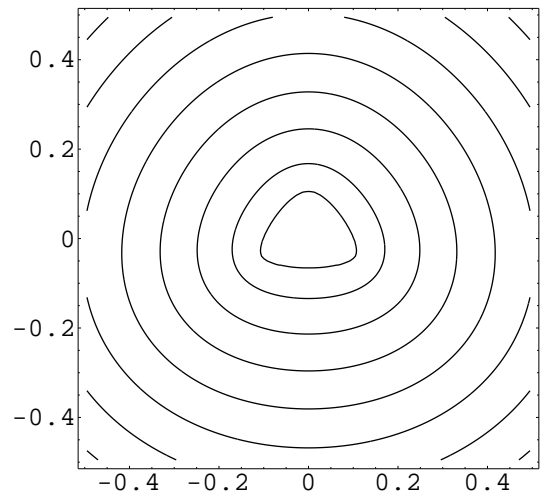}\\

\hspace{3cm}\psfig{file=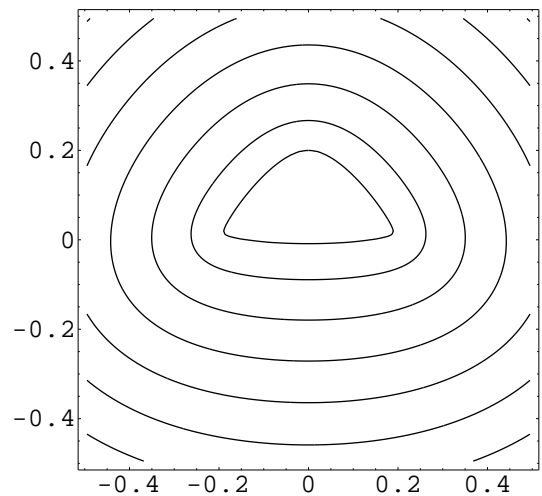}\\

\caption{ Level sets of the initial condition (a) and an intermediate 
stage before the collapse, when the averaged radius decreased by a factor
$2^6$ (b) and $2^{9}$ (c). This flow is not self-similar.} \label{nonself}
\end{figure}

In Fig. \ref{nonself}-a we consider the following example with initial
condition 
\begin{equation}
v^{(0)}(r,\theta ,t_0)=r-a_0\left( 1-\varepsilon \cos 3\theta -
\varepsilon \cos 5\theta \right)
,\qquad \varepsilon =0.05  \label{initial2}
\end{equation}
Clearly, this shape does not have k-fold symmetry for any $k>2$. This combination
of
modes yields a slightly elongated figure. 
The evolution for $m=1.5$ leads to the intermediate shapes shown in Fig.
\ref{nonself}(b)-(c). Initially the 3-fold 
component of the initial condition dominates and the shape is almost triangular,
but, since the initial
shape was slighty ellongated in the $x$
direction, the final shape becomes increasingly ellongated. As far as the
computation shows, the solution
does not reach a self-similar regime in this case.
The shape of the hole becomes increasingly elongated, until the numerical grid 
cannot accurately resolve the shape 
of the interface. We can interpret this result in terms of the interaction of the
modes: the new mode that accounts for the ellongation, namely $k=2$, is created
from the
non-linear interaction of the modes $3$ and
$5$. 

If instead of the modes $3$ and $5$ we start with the modes $3$ and $6$, the shape
will
have 3-fold symmetry, and a self-similar solution for the most dominant
mode ($k=3$)
is eventually reached.
Further discussions of these cases will appear in ref. \cite{Betelu99}.\\
\\ 
{\bf ACKNOWLEDGMENTS}
S. I. B. is grateful to the School of Mathematics of the University of 
Minnesota
for a Visiting Position in 1999. We also thank to John Lowengrub
for
his help
in the numerical implementation of the ENO method, Roberto Fernandez for
bringing to our attention the reference \cite{Golden} and its relationship
with
diffusion, and to Nigel Goldenfeld for the reference \cite{Chen}.

\section{Appendix 1}

In order to compute the $\left| \nabla v\right| ^2$ term, we use the ENO
scheme, which has been used very succesfully in the numerical solution of
Hamilton-Jacobi equations\cite{Osher91,Shu88}. The ENO scheme is an adaptive
stencil interpolation procedure which automatically obtains information from
the locally smoothest region, and hence yields a uniform high essentially
nonoscillatory approximation for piecewise smooth functions.

When computing the partial derivative respect to $r$ corresponding to $%
v_{jl} $ at discrete nodes $r_j=hj$, $j=0,\pm 1,\pm 2,...$ we first write
the undivided differences 
\[
w(j,0)=v_{jl} ,\hspace{1.5cm}  
\]
\[
w(j,k)=w(j+1,k-1)-w(j,k-1) \hspace{1.5cm} k=1,...,d+1 
\]
where $d$ is the order of the approximation ($d=2$ in our case).
The ENO stencil-choosing procedure is optimally implemented by starting with 
$i(j)=j$ and performing 
\[
if\qquad \left| w(i(j),k)\right| >\left| w(i(j)-1,k)\right| \qquad
then\qquad i(j)= i(j)-1 
\]
for $k=2,...,d$. Finally we compute the forward and the backward
derivatives, 
\[
\left( \partial v/\partial r\right) ^{+}=\frac
1h\sum_{k=1}^dc(i(j)-j),k)w(i(j),k) 
\]
and 
\[
\left( \partial v/\partial r\right) ^{-}=\frac
1h\sum_{k=1}^dc(i(j-1)-j),k)w(i(j-1),k) 
\]
where 
\[
c(m,k)=\frac 1{k!}{\sum_{s=m}^{m+k-1}} \prod_{l=m,l\neq s}^{m+k-1}(-l). 
\]
The same procedure is used to compute the derivatives in the $\theta $
direction. In order to compute the upwind approximation for $v_r^2$ we write 
\[
v_r^2=min\left( \left( \partial v/\partial r\right) ^{-},0\right)
^2+max\left( \left( \partial v/\partial r\right) ^{+},0\right) ^2. 
\]
Similar expressions are used to compute $v_\theta ^2$. This scheme is very
useful when there are discontinuities in the first derivatives, as in our
case at the interface and for $m=1$.

\section{Appendix 2}

In this Appendix we describe the linearized stability analysis of the
Graveleau interfaces in dimension $d=2$. Although the basic equations are
derived and analyzed in \cite{Ange97}, there remain several questions which, at
least for the present, can only be answered by numerical studies. We
describe these studies here, and for the convenience of the reader give a
brief outline of the derivation of the relevant equations.

We begin by rewriting the porous medium pressure Eq. (\ref{pressure}) in
self-similar coordinates. Let $v({\bf x},t)$ be a solution of Eq. (\ref{pressure})
and
set
\[
v({\bf x},t)=(T-t)^{2\beta -1}V({\bf y},\tau ),
\]
where
\[
{\bf y}=\frac {\bf x}{(T-t)^\beta }\hspace{1cm}\mbox{ and }\hspace{1cm} \tau =-\ln
(T-t).
\]
Then $V$ satisfies the equation
\begin{equation}
\frac{\partial V}{\partial \tau }=nV\Delta V+\left| \nabla V\right|
^2-\beta
{\bf y}\cdot \nabla V+(2\beta -1)V,  \label{A.1}
\end{equation}
where $n=m-1$. Let $\rho =\left| {\bf y}\right| $ and write the Graveleau
solution
(Eq. (\ref{3})) normalized with $c=1$ in the form
\[
v_1({\bf x},t)=(T-t)^{2\beta -1}\psi \left( \rho \right) .
\]
Note that $\psi $ is a steady state solution to Eq. (\ref{A.1}), i.e.,
\[
n\psi \left( \psi ^{\prime \prime }+\frac{d-1}\rho \psi ^{\prime }\right)
+\left( \psi ^{\prime }\right) ^2-\beta \rho \psi ^{\prime }+(2\beta
-1)\psi
=0.
\]
Moreover,
\[
\psi \left( \rho \right) \left\{
\begin{tabular}{l}
$=0$ for $0\leq \rho \leq \gamma ^{-\beta }$ \\
$>0$ for $\rho >\gamma ^{-\beta }$%
\end{tabular}
\right. ,
\]
and
\[
\psi \left( \rho \right) \sim \rho ^{2-\frac 1\beta }\mbox{ as }\rho
\rightarrow \infty
\]
(cf. \cite{Aro93}).
We now restrict our attention to flows in two dimensions and introduce
polar
coordinates
\[
{\bf y}=\rho \left( \cos \theta ,\sin \theta \right) ,
\]
where
\[
\rho =\sqrt{y_1^2+y_2^2}\mbox{ and }\theta =\arctan \frac{y_2}{y_1}.
\]
If we set $W(\rho ,\theta ,\tau )=V({\bf y},\tau )$ the $W$ satisfies
\[
\frac{\partial W}{\partial \tau }=nW\left( \frac{\partial ^2W}{\partial
\rho
^2}+\frac 1\rho \frac{\partial W}{\partial \rho }+\frac 1{\rho ^2}\frac{%
\partial ^2W}{\partial \theta ^2}\right) +\left( \frac{\partial
W}{\partial
\rho }\right) ^2+\left( \frac 1\rho \frac{\partial W}{\partial \theta }%
\right) ^2
\]
\begin{equation}
-\beta \rho \frac{\partial W}{\partial \rho }+\left( 2\beta -1\right) W.
\label{A.2}
\end{equation}

For $p\geq 0$ the level curves of $W$ are given implicitly by
\[
W(\rho ,\theta ,\tau )=p.
\]
Assuming that $\frac{\partial W}{\partial \rho }\neq 0$ we can solve for
$%
\rho $ to get
\[ 
\rho =S(p,\theta ,\tau ).
\]
All of the derivatives of $W$ which appear in equation (\ref{A.2}) can be
calculated from the relation
\[
W(S(p,\theta ,\tau ),\theta ,\tau )-p=0
\]
(cf. \cite{Ange97} for details). Thus we derive the evolution equation for the
level
curves $\rho =S$:
\[
\left( S\frac{\partial S}{\partial p}\right) ^2\frac{\partial S}{\partial
\tau }=np\left\{ S^2\frac{\partial ^2S}{\partial p^2}-S\left(
\frac{\partial
S}{\partial p}\right) ^2+\frac{\partial ^2S}{\partial p^2}\left( \frac{%
\partial S}{\partial \theta }\right) ^2-2\frac{\partial S}{\partial
p}\frac{%
\partial S}{\partial \theta }\frac{\partial ^2S}{\partial p\partial \theta
}%
\right.
\]
\begin{equation}
\left. +\frac{\partial ^2S}{\partial ^2\theta }\left( \frac{\partial S}{%
\partial p}\right) ^2\right\} +\beta S^3\left( \frac{\partial S}{\partial
p}%
\right) ^2-S^2\frac{\partial S}{\partial p}-\frac{\partial S}{\partial p}%
\left( \frac{\partial S}{\partial \theta }\right) ^2-(2\beta -1)pS^2\left(
\frac{\partial S}{\partial p}\right) ^3.  \label{A.3}
\end{equation}  

The Graveleau function $\psi (\rho )$ is an increasing function for $\rho
>\gamma ^{-\beta }$ with range $[0,\infty )$. Thus we can invert to obtain
$%
\rho =\Psi (p)$ which is an increasing function for $p\geq 0$ with range
$%
[\gamma ^{-\beta },\infty ).$ Note that $\Psi $ is a steady state solution
to Eq. (\ref{A.3}) and satisfies
\begin{equation}
np\Psi \Psi ^{\prime \prime }=np(\Psi ^{\prime })^2-\beta (\Psi \Psi
^{\prime })^2+\Psi \Psi ^{\prime }+(2\beta -1)p\Psi (\Psi ^{\prime })^3,   
\label{A.}
\end{equation}
where $^{\prime }=d/dp$. 
Set $S(p,\theta ,\tau )=\Psi (p)+\xi (p,\theta ,\tau )$, where we assume
that $\left| \xi \right| \ll 1.$ \newpage To leading order $\xi $ satisfies the
linear equation
\[
(\Psi \Psi ^{\prime })^2\frac{\partial \xi }{\partial \tau }=np\Psi
^2\frac{%
\partial ^2\xi }{\partial p^2}+np\left( \Psi ^{\prime }\right) ^2\frac{%
\partial ^2\xi }{\partial \theta ^2}
\]
\begin{equation}
+\left\{ 2\beta \Psi ^3\Psi ^{\prime }-\Psi ^2-2np\Psi \Psi ^{\prime
}-3(2\beta -1)p(\Psi \Psi ^{\prime })^2\right\} \frac{\partial \xi }{%
\partial p}  \label{A.4}
\end{equation}
\[
+(np+\beta \Psi ^2)(\Psi ^{\prime })^2\xi .
\]
Since the coefficients in Eq. (\ref{A.4}) depend only on $p$ there are
solutions of
the
form
\[
\xi (p,\theta ,\tau )=A(p)\exp \left( ik\theta +\lambda \tau \right) ,
\]
where $A$ satisfies the ordinary differential equation
\begin{equation}
np\Psi ^2A^{\prime \prime }+\left\{ 2\beta \Psi ^{\prime }\Psi ^3-\Psi   
^2-2np\Psi \Psi ^{\prime }-3(2\beta -1)p(\Psi \Psi ^{\prime })^2\right\}
A^{\prime }  \label{A.5}
\end{equation}
\[
+\left\{ np(1-k^2)+(\beta -\lambda )\Psi ^2\right\} (\Psi ^{\prime
})^2A=0.
\]
Here $k$ is the given wave number and $\lambda $ is an eigenvalue which
must
be determined.
Since $\Psi (0)=\gamma ^{-\beta }$ it follows from equation \ref{A.3} that
$\Psi
^{\prime }(0)=\gamma ^\beta /\beta $. Thus, to leading order for $p\ll 1$,
equation \ref{A.5} becomes
\[
npA^{\prime \prime }+A^{\prime }+(\beta -\lambda )\gamma ^{2\beta }A=0.
\]
This equation has a regular singular point at $p=0$, and possesses a
unique  
analytic solution determined by the initial conditions
\begin{equation}
A(0)=1\mbox{ and }A^{\prime }(0)=\frac{\lambda -\beta }\beta \gamma
^{2\beta
}.  \label{A.6}   
\end{equation}

For $p\gg 1$, we have 
\[
\Psi \sim p^{\frac \beta {2\beta -1}}\mbox{ and }\Psi ^{\prime }\sim \frac
\beta {2\beta -1}p^{\frac{1-\beta }{2\beta -1}}.
\]
Thus, to leading order for $p\gg 1$, equation \ref{A.5} becomes
\[  
np^{\frac{4\beta -3}{2\beta -1}}A^{\prime \prime }-\frac{\beta ^2p}{2\beta
-1%
}A^{\prime }+\frac{\beta ^2(\beta -\lambda )}{(2\beta -1)^2}A=0.
\]
Most solutions of this equation grow exponentially at infinity with
\[
\ln A^{\prime }\sim \frac{\beta ^2}np^{\frac 1{2\beta -1}},
\]
but there are also solutions with algebraic growth
\begin{equation}
A=O\left( p^{\frac{\beta -\lambda }{2\beta -1}}\right) .  \label{A.7}
\end{equation}
The eigenvalues are those values of $\lambda $ for which the solution of
Eq. (\ref{A.5}) with initial values of Eq. (\ref{A.6}) has the algebraic
growth given
by Eq. (\ref{A.7}) at  
infinity. They are obtained numerically by a shooting technique.

The eigenvalues of equation (\ref{A.5}) are analyzed in detail in
reference
\cite{Ange97},
where it is shown that for each $m\in (1,\infty )$ they form a doubly
infinite sequence $\lambda _{kj}(m)$. Here $k\geq 0$ is the wave number
and $%
j\geq 0$ is the number of zeros of the corresponding eigenfunction.
Moreover, $\lambda _{01}(m)=0$, $\lambda _{0j}(m)<0$ for $j\geq 2$,
$\lambda
_{kj}(m)<0$ for $k\geq 1$ and $j\geq 1$, and $\lambda _{k0}(m)>0$ for
$k=0$
or $1$. Bifurcations occur for those values of $m$ where $\lambda
_{kj}(m)$  
changes sign, and this can only happen for $j=0$ and $k\geq 2$. In \cite{Ange97}
it
is shown that there are infinitely many bifurcations. Specifically, for
every $m^{\prime }\in (1,\infty )$ there exists an integer
$k_{*}(m^{\prime 
})\geq 2$ such that for each $k>k_{*}(m^{\prime })$ there is a bifurcation
point $m_k\in (1,m^{\prime })$. The bifurcating solutions are not radially
symmetric, but do possess the $k$-fold symmetry of $\cos (k\theta )$.

Numerical studies of the eigenvalues indicate that there are no
bifurcations
for wave number $k=2$ for any value of $m$, and that there is a unique
bifurcation point $m_k$ for each $k\geq 3$. In particular,
\[
\lambda _{k0}(m)\left\{
\begin{tabular}{l}
$>0$ for $m\in (1,m_k)$ \\
$<0$ for $m\in (m_k,\infty )$%
\end{tabular}
\right. .
\]
The $m_k$ are ordered with
\[
\infty >m_3>m_4>\cdot \cdot \cdot >m_k>\cdot \cdot \cdot \searrow 1.
\]
Table IV summarizes some of the numerical results for the sign of $\lambda
_{k0}(m).$ A minus sign in the position $(k,m)$ indicates stability with
respect to perturbations with wave number $k$ for the given value of $m$,
while a plus sign indicates instability. The bifurcation points $m_k$
occur 
between adjacent $m$-values where the $\lambda _{k0}(m)$ have opposite
signs. Thus, for example, $m_4\in (1.320,1.321)$.\newpage
\[
\begin{tabular}{lllllllllll}
& $m=$ & $1.12$ & $1.13$ & $1.18$ & $1.19$ & $1.320$ & $1.321$ & $1.69$ &
$%
1.7$ & $2$ \\
$k=$ & $2$ & $+$ & $+$ & $+$ & $+$ & $+$ & $+$ & $+$ & $+$ & $+$ \\
& $3$ & $+$ & $+$ & $+$ & $+$ & $+$ & $+$ & $+$ & $-$ & $-$ \\
& $4$ & $+$ & $+$ & $+$ & $+$ & $+$ & $-$ & $-$ & $-$ & $-$ \\
& $5$ & $+$ & $+$ & $+$ & $-$ & $-$ & $-$ & $-$ & $-$ & $-$ \\
& $6$ & $+$ & $-$ & $-$ & $-$ & $-$ & $-$ & $-$ & $-$ & $-$%
\end{tabular}
\]
\begin{center}
\textbf{Table IV:} Sign of $\lambda_{k0}(m)$. Negative values indicate
that the circular self-similar solution is stable for the corresponding
pair of $m$ and $k$.   
\end{center}

\pagestyle{empty}
\begin{center}
\begin{figure}
\psfig{file=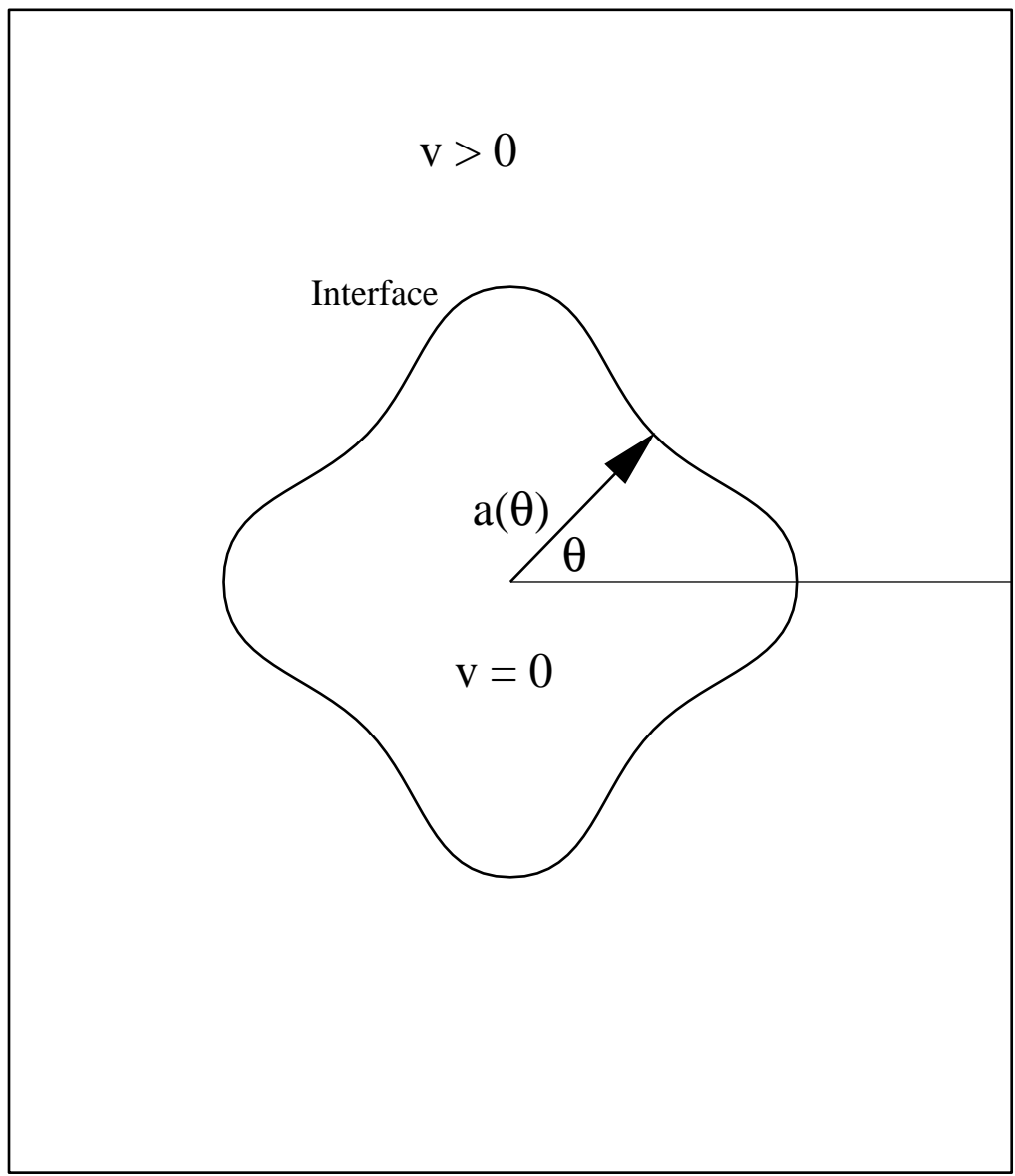}
\caption{ Sketch of a noncircular convergent flow. The fluid lies
outside the interface and flows inwards.}\label{sketch}
\end{figure}

\begin{figure}
\psfig{file=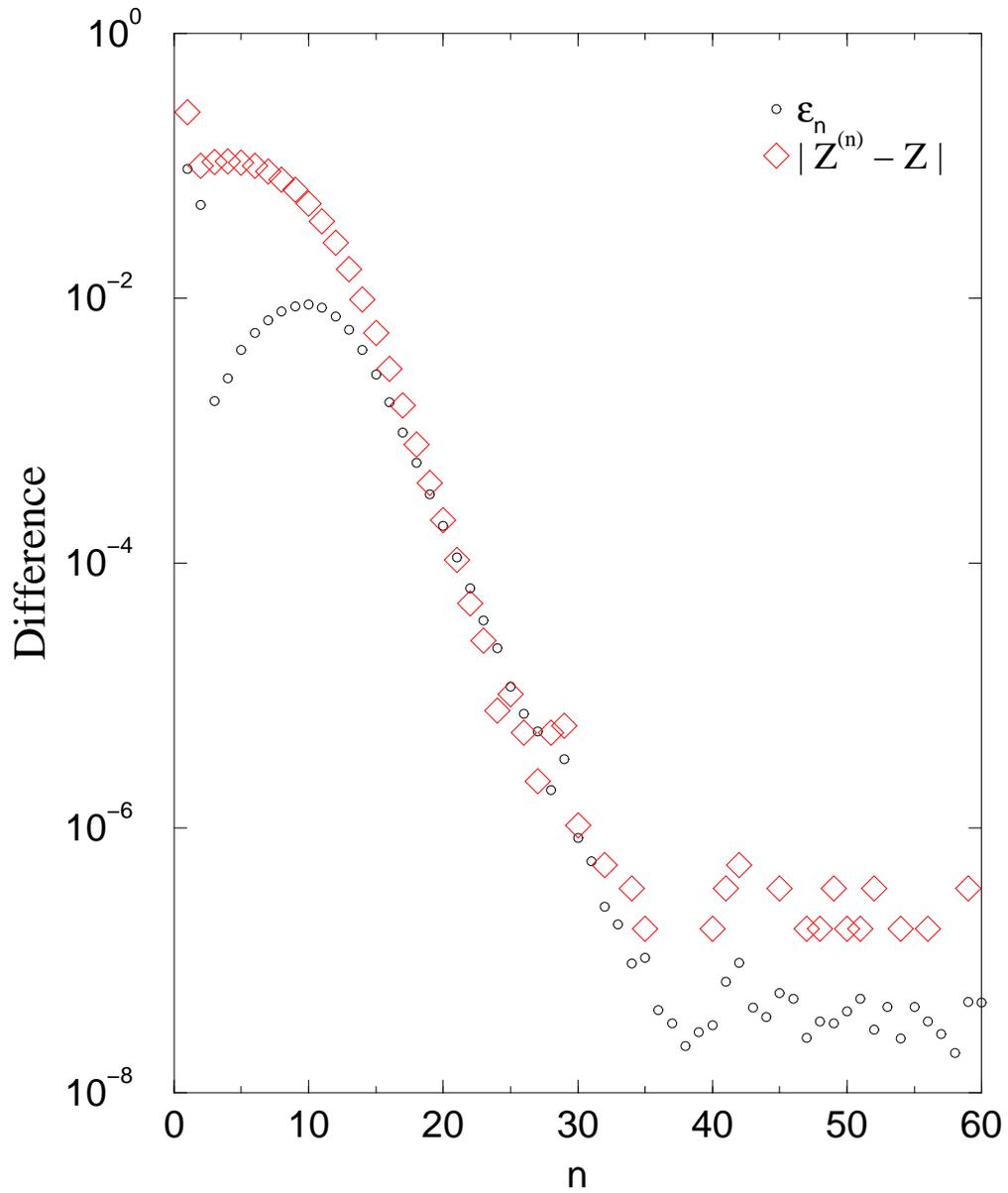}
\caption{ Convergence of the numerical solution: the difference between
succesive approximations tend to zero and the renormalization constant $Z$
tends to a constant as the number of renormalizations $n$ increases. In this case
$m=1.5$, $N_r=300$, $N_\theta=300$ and the initial shape is a
perturbated circle with $k=3$.} \label{convergence}
\end{figure}
\end{center}

\flushright

\center

\end{document}